\documentclass{mn2e}

\usepackage{amssymb}
\usepackage{amsmath}
\usepackage{epsfig}
\title{The Quasar Mass-Luminosity Plane III: Smaller Errors on Virial Mass Estimates}
\author[Charles L. Steinhardt and Martin Elvis]
       {Charles L. Steinhardt and Martin Elvis \\
        Harvard-Smithsonian Center for Astrophysics, 60 Garden St, Cambridge, MA
	02138} 
\date{\today}
\begin{document}
\maketitle

\label{firstpage}

\begin{abstract}
We use 62,185 quasars from the Sloan Digital Sky Survey DR5 sample to explore
the quasar mass-luminosity plane view of virial mass estimation.  Previous work
shows deviations of $\sim 0.4$~dex between virial and reverberation masses.  The
decline in quasar number density for the highest-Eddington ratio quasars at each
redshift provides an upper bound of between $0.13$ and $0.29$ dex for virial mass
estimates.  Across different redshift bins, the maximum possible Mg{\small II}
mass uncertainties average $0.15$ dex, while H$\beta$ uncertainties average
$0.21$ dex and C{\small IV} uncertainties average $0.27$ dex.  Any physical
spread near the high-Eddington-ratio boundary will produce a more restrictive
bound.  A comparison of the sub-Eddington boundary slope using H$\beta$ and Mg{\small II} masses finds better agreement with uncorrected Mg{\small II} masses than with recently proposed corrections.  The best agreement for these bright objects is produced by a multiplicative correction by a factor of 1.19, smaller than the factor of 1.8 previously reported as producing the best agreement for the entire SDSS sample.
\end{abstract}


\section{Introduction}
\label{sec:intro}

Strong correlations between the mass of supermassive black holes (SMBH) and the
stellar velocity dispersion \cite{msigma1,msigma2} and luminosity
\cite{FerrareseFord2005} of their host galaxies argue that our understanding of
galactic formation is incomplete without an understanding of the SMBH found at
their centres.  Our modern arsenal for learning about SMBH growth is predicated
on four basic tools: (1) Large samples containing $\sim 10^5$ quasars provided
by modern surveys \cite{Schneider2007,2MASS,2QZ}; (2) Bolometric luminosity
estimation comparing a piece of the spectrum \cite{Richards2006b} to templates
made from composite quasar spectra \cite{Elvis1994}; (3) The Soltan argument
\cite{Soltan1982,Salucci1999,Yu2002} that the integrated luminosity density of
active galactic nuclei is consistent with the mass density in the local SMBH
population; and (4) A series of SMBH mass estimation techniques comprising the
``BH Mass Ladder'' \cite{Peterson2004}.   

Until recently, the first three of these tools might have been thought
sufficient.  The highest-redshift rung of the black hole ``ladder'', virial
masses, is the only option at redshift $z \gtrsim 0.2$, and comparison with
reverberation mapping yields a $\sim 0.4$~dex statistical uncertainty
\cite{Vestergaard2006}.  Further, early constructions of the quasar
mass-luminosity distribution showed a population of quasars all within $\sim 1$
dex of their Eddington luminosity \cite{Kollmeier2006}. 

If quasars are all within $\sim 1$ dex of their Eddington luminosity
\cite{Kollmeier2006}, then quasar luminosity would be a good proxy for
supermassive black hole (SMBH) mass.  Quasars could then be modelled as
``light-bulbs'' of a characteristic wattage, either operating near their
Eddington luminosity or lying dormant (although Hopkins \& Hernquist
2008\nocite{Hopkins2008} argue this is an oversimplification).  If the
relationship between quasar mass and luminosity were truly this simple, virial
mass estimation would have little to add to the information already contained in
quasar luminosity functions predicated on higher-precision bolometric luminosity
estimates.

In Papers I and II \cite{Steinhardt2009a,Steinhardt2009b}, we discovered that
the quasar mass-luminosity plane is quite complex. We demonstrated the existence
of several surprising new boundaries in the mass-luminosity plane.  So, quasar
mass and luminosity functions are not enough; these projections of the quasar
mass-luminosity plane hide information.  In Paper II \nocite{Steinhardt2009b},
we showed that the ratio of emission lines from the broad line region may be
evolving as quasars approach turnoff.  High-precision black hole mass estimation
would allow the identification of specific quasars very close to these
boundaries as part of a search for a signature of their underlying physical
causes.  However, SMBH masses cannot simply be well-estimated from the quasar
luminosity, and so the investigation of SMBH evolution indeed requires all
four tools.

Our initial work relied upon H$\beta$- and C{\small IV}-based virial mass
estimates from Vestergaard \& Peterson (2006)\nocite{Vestergaard2006} and
Mg{\small II}-based estimators from McLure \& Dunlop (2004)\nocite{McLure2004}.
There have been several attempts to improve these virial mass scaling
relationships \cite{Onken2008,Risaliti2009,Marconi2009}.  In this paper, we
reconsider virial mass scaling relations in light of these new boundaries in the quasar $M-L$ plane.

Every boundary in the SDSS quasar locus in the mass-luminosity plane is a
combination of a real, underlying physical boundary, SDSS selection, statistical uncertainty, and systematic uncertainty.  The Eddington luminosity and
sub-Eddington boundary (SEB) characterise the quasars with the highest Eddington
ratios at each mass, or equivalently, quasars with the highest Eddington ratio
at fixed luminosity.  While we have termed this boundary ``sub-Eddington''
because of its behaviour at high mass, the SEB is flatter than the Eddington
limit and for the lowest-mass objects at most redshift, the Eddington luminosity
is the more restrictive bound.  As discussed in Paper I\nocite{Steinhardt2009a},
the SDSS selection function is not a factor in the location of the SEB.
Therefore, the decline in number density is a combination of statistical
uncertainty, systematic uncertainty, and an underlying physical cutoff, and as
such, its sharpness can be used to derive an upper bound on the maximum
statistical uncertainty in virial mass estimation.  The statistical uncertainty
has been estimated as $0.4$~dex via comparison of virial and reverberation
masses \cite{Vestergaard2006}.

In \S~\ref{sec:error}, we use the sharpness of boundaries in mass to place a
tighter upper bound on the statistical uncertainty of virial mass estimates.  In
\S~\ref{sec:correction}, we show that proposed corrections to Mg{\small
II}-based virial mass estimation are inconsistent for the brightest quasars at
$0.6 < z < 0.8$.  We discuss the implications of these results in
\S~\ref{sec:discussion}.

\section{A New Bound on Statistical Uncertainties in SMBH Masses}
\label{sec:error}

\begin{figure}
  \epsfxsize=3in\epsfbox{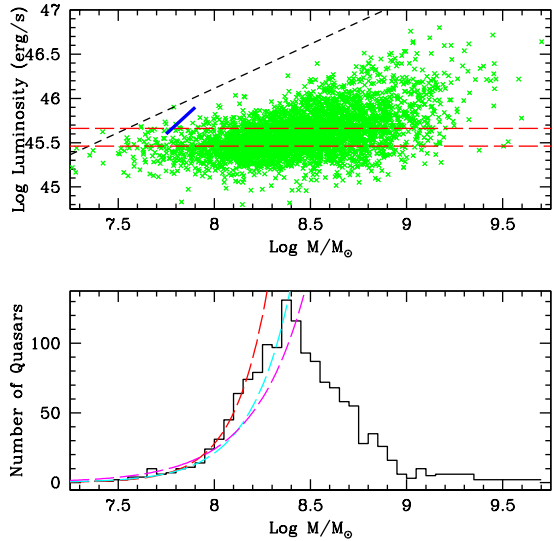}
\caption{Top: The quasar $M-L$ plane for Mg{\small II} masses at $0.4 < z < 0.6$.  We take the quasar number density as a function of mass around the average luminosity (between the red, dashed lines).  Quasar luminosity variability of $0.3$~dex  might lead to a change in estimated virial mass of $0.15$~dex (blue).  Bottom: The low-mass decay in quasar number density as a function of mass at $45.46 < \log L_{bol} < 45.66$ is best-fitting by an exponential decay with an e-folding of $0.15$~dex (red).  For comparison, the best-fitting decays using the average e-foldings for H$\beta$ (cyan, $0.21$~dex) and C{\small IV} (magenta, $0.27$~dex) are also indicated.}
\label{fig:decays}
\end{figure}

We use black hole masses for 62,185 of the 77,429 SDSS DR5 quasars
\cite{Schneider2007} as determined by Shen et al. (2008)\nocite{Shen2008} using
H$\beta$- and C{\small IV}-based virial mass estimators from Vestergaard \&
Peterson (2006)\nocite{Vestergaard2006} and Mg{\small II}-based estimators from
McLure \& Dunlop (2004)\nocite{McLure2004}.  We divide the SDSS quasar
population into 14 redshift bins, of width 0.2 below $z = 2$ and wider at higher
redshift.  In each bin, we consider the mass distribution of quasars in a luminosity
bin of width 0.2 dex centred at the average bolometric luminosity for the
catalogue at that redshift.  Choosing this fixed luminosity gives us a large
number of objects in our attempt to fit the decay rate at low mass.  The decay at low mass is more rapid than at high mass.  In each
bin, we fit the decline in number density as both a Gaussian and an exponential
decay (Figure \ref{fig:decays}), reporting the dispersion or e-folding in Table
\ref{table:decayfits}.  

\begin{table}
\caption{Best-fitting forms for the decay in quasar number density as a function of
mass.  These e-foldings imply maximum statistical uncertainties of, on average,
$0.21$ dex for H$\beta$-based virial masses, $0.15$ dex for Mg{\small II}, and
$0.27$ dex for C{\small IV}.}
\begin{tabular}{|c|c|c|c|c|c|c|}
\hline 
$z$ & $N$ & $\left<\log L\right>$ & $\sigma$ (dex) & $\chi^2_\nu$ & e-folding (dex) & $\chi^2_\nu$ \\
\hline 
H$\beta$ & & & & & \\
0.2-0.4 & 2690 & 45.25 & 0.23 & 2.93 & 0.19 & 0.77 \\
0.4-0.6 & 4250 & 45.54 & 0.18 & 2.73 & 0.20 & 0.75 \\
0.6-0.8 & 3665 & 45.89 & 0.24 & 1.40 & 0.24 & 0.90 \\
Mg{\small II} & & & & & \\
0.4-0.6 & 4203 & 45.56 & 0.16 & 0.71 & 0.17 & 0.58 \\
0.6-0.8 & 4727 & 45.80 & 0.19 & 2.80 & 0.16 & 1.00 \\
0.8-1.0 & 5197 & 46.02 & 0.13 & 3.58 & 0.15 & 1.24 \\
1.0-1.2 & 6054 & 46.21 & 0.22 & 1.55 & 0.13 & 1.05 \\
1.2-1.4 & 7005 & 46.32 & 0.14 & 3.18 & 0.14 & 1.50 \\
1.4-1.6 & 7513 & 46.43 & 0.18 & 2.22 & 0.13 & 1.10 \\
1.6-1.8 & 6639 & 46.57 & 0.23 & 2.29 & 0.18 & 0.53 \\
1.8-2.0 & 4900 & 46.71 & 0.27 & 1.32 & 0.19 & 0.96 \\
C{\small IV} & & & & & \\
1.8-2.0 & 4627 & 46.60 & 0.25 & 2.68 & 0.26 & 0.71 \\
2.0-3.0 & 7079 & 46.79 & 0.27 & 4.34 & 0.27 & 0.94 \\
3.0-4.1 & 2859 & 46.98 & 0.29 & 3.64 & 0.29 & 1.38 \\
\hline  
\end{tabular}
\label{table:decayfits}
\end{table}

What form should we expect from the decay in quasar number density?  Let the
true quasar mass distribution be $\rho_p(M)$ and let $\phi(x)$ be the
probability distribution that a virial mass is incorrect by $x$.  Then, the
observed quasar mass distribution $\rho_o(M)$ is the convolution
\begin{equation}
\rho_o(M) = (\rho_p \ast \phi)(M) = \int_{-\infty}^{\infty} \rho_p(\mu) \phi(M-\mu) d\mu.
\end{equation}  
For example, if $\rho_p(M)$ were a step (Heaviside) function, i.e., a constant
number density at low mass and zero above the SEB, and the uncertainty $\phi(x)$
were Gaussian, we would see a decay taking the form of the error function, $1 +
Erf(x) = \int_{-\infty}^x e^{-t^2} dt.$ In practice, the exact form of the tail
is highly sensitive to $\rho_p$, and the exponential decay described in Table
\ref{table:decayfits} is a better fit to the low-mass tail of $\rho_o$ than a
Gaussian (Table \ref{table:decayfits}), polynomial, or $Erf(x)$.  The exact form
of $\rho_p$ appears to be more complicated than a step function at masses above
the SEB (Figure \ref{fig:decays}).  However, the convolution acts to spread out
the signal, and therefore the dispersion of features in $\rho_o$ will be at
least as large as those in $\phi(x)$.  So, the e-foldings of best-fitting
exponential decays in Table \ref{table:decayfits} are inconsistent with an 0.4
dex statistical uncertainty for virial mass estimation.

\subsection{Effects of quasar variability on virial mass estimation}

Virial mass estimates take the form
\begin{eqnarray}
\label{vmeform}
\log (M/M_\odot) & = & A \\ & + & \log
\left[\left(\frac{\textrm{FWHM}(\textrm{BLR line})}{1000\textrm{
km/s}}\right)^2\left(\frac{\lambda L_\lambda (B~{\rm
\AA})}{10^{44}\textrm{ erg/s}}\right)^{C}\right] \nonumber,
\end{eqnarray}
for emission lines in the broad line region (BLR) and a nearby continuum flux.
For the virial mass estimates used in the Shen et al. (2008)\nocite{Shen2008}
catalogue, $C$ is between $0.50$ and $0.53$
\cite{McLure2002,McLure2004,Vestergaard2006}.  One possible explanation for this
discrepancy, then, is that the analysis above only considers quasars at fixed
bolometric luminosity, while the adjacent continuum is one component of virial
mass estimation.  Perhaps the statistical uncertainty in virial mass estimation
is mostly caused by quasar variability, in which the bolometric luminosity (and
adjacent continuum) change on a timescale of years while the black hole mass
remains very nearly constant.  

Variability moves objects along a $L = M^2$ line (Figure \ref{fig:decays}), blurring each thin luminosity slice along the mass axis by 0.18~dex for typical long-term, 0.3~dex optical variations \cite{deVries2005}.  This blurring is larger than most of the Mg{\small II} decays, notably smaller than the C{\small IV} decays, but is consistent with the H$\beta$ decays. The additional  C{\small IV} scatter must have some other cause.
Variability is wavelength dependent in quasars, being stronger toward the UV,
but the difference between 2800 \AA~and 1500 \AA~is too small to explain the
different Mg{\small II} and  C{\small IV} decays

Similarly, the adjacent luminosity used to estimate the black hole mass is based on the continuum local to
each line, while the bolometric luminosity is based on the five SDSS photometry
points.  Thus, systematic offsets between these two luminosities could cause a
scatter. The correction would have to have a peculiar shape as the longest and
shortest H$\beta$ and C{\small IV} measurements both have larger scatter than
the intermediate wavelength Mg{\small II} value.

\section{The Mg{\small II} View of the Sub-Eddington Boundary}
\label{sec:correction}

Mg{\small II} has the sharpest boundary in the $M-L$ plane, and so appears to be the most precise mass indicator. Currently, though, H$\beta$-based masses are
considered to be the most reliable, because they have been calibrated directly
against reverberation masses at low redshift. Certainly, several potential problems with C{\small IV} virial masses have
been suggested \cite{Shen2008,Marconi2009}.

Corrections have also been suggested to the Mg{\small II} masses derived by
McLure \& Dunlop (2004)\nocite{McLure2004} (MD04).  Onken \& Kollmeier
(2008)\nocite{Onken2008} examine SMBH for which both H$\beta$ and Mg{\small II}
masses are available and find that the Mg{\small II}-based $M_{BH}$ may be
overestimated at high Eddington ratio and underestimated at low Eddington ratio.
Risaliti, Young, \& Elvis (2009)\nocite{Risaliti2009} (RYE09) quantify this
correction empirically as
\begin{equation}
\log[M_{BH}(H\beta)]=1.8\times\log[M_{BH}(Mg{\small II})]-6.8.
\end{equation}
In addition to correcting the central values of Mg{\small II} virial masses, the
RYE09 correction also increases the e-folding decay by the same factor
of 1.8, to an average of $0.27$ dex, which would make Mg{\small II} masses less
precise than H$\beta$ and comparable to C{\small IV}.

This multiplicative correction is surprising, because it requires that the gas
emitting either H$\beta$ or Mg{\small II} lines has a non-virial component.  If
H$\beta$ is virial, as expected from calibration between H$\beta$ and
reverberation masses \cite{Vestergaard2006}, then the mass-velocity relation for
Mg{\small II} would need to be $M \propto v^{3.6}$.  RYE09 propose that this mismatch might instead be due to uncertainties in
measurement of the Mg{\small II} line, mainly because of potential Fe{\small II}
contamination.  

\begin{figure}
  \epsfxsize=3in\epsfbox{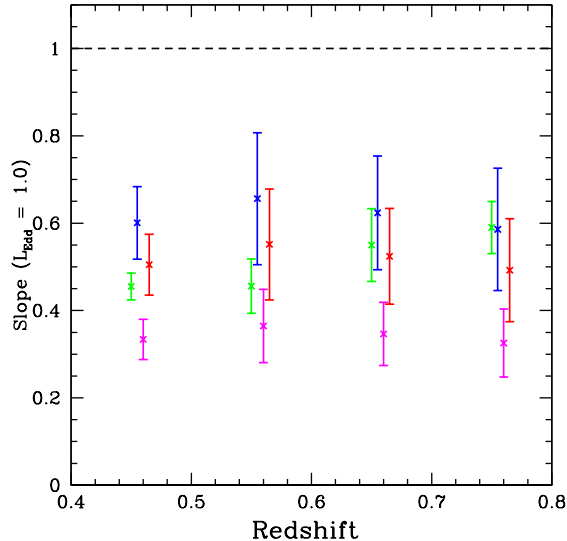}
\caption{A comparison of best-fitting slopes to the sub-Eddington boundary in
four redshift ranges using different virial mass scaling relations.  The
H$\beta$-based estimates (green) are a better fit for the original McLure \&
Dunlop (2004) Mg{\small II} masses (blue) than the Risaliti, Young, \& Elvis
(2009) correction (magenta).  The best match between slopes is produced by a
smaller Mg{\small II} correction (red).  The black dashed line is drawn at the slope of the Eddington luminosity}
\label{fig:sebslopes}
\end{figure}

To test the multiplicative correction we examine the SEB more closely.  
The slope of the SEB is sensitive to the multiplicative correction to Mg{\small II}-based virial masses but not to any additive correction. As the
SEB is composed of the brightest quasars in each redshift bin they typically
have the best-measured spectra at each mass, minimising Mg{\small II} measurement errors.
Both Mg{\small II} and H$\beta$ masses are available at the $0.4 < z < 0.8$.
Using the techniques detailed in Paper I we subdivided this bin into four
redshift bins of width $\Delta$z= 0.1. We then calculated the slope of the
best-fitting SEB in each bin using both the MD04 scaling relation and the RYE09
correction.    In
Figure \ref{fig:sebslopes}, we show the four H$\beta$ SEB slopes (green) compared to
slopes using MD04 (blue) and RYE09 (magenta).

In each redshift bin, the deviations between MD04 and H$\beta$ slopes are
between 0.3 $\sigma$ and 1.6 $\sigma$ (MD04 averages $0.8 \sigma$ higher).
Deviations between RYE09 and H$\beta$ are between 0.9 and 2.7 $\sigma$ (average
1.9 $\sigma$), with the RYE09 slope always lower.  If Mg{\small II} masses with the RYE08 correction produced identical SEB slope estimates to H$\beta$, the probability that all four measurements would be this far below the H$\beta$ slope is 0.2\% (comparing to a Monte Carlo of normally-distributed measurements).
A best-fitting correction between H$\beta$ and Mg{\small II} slopes
(Figure \ref{fig:sebslopes}) reduces the factor of 1.8 to one of 1.19.  
The $M-L$ plane view of Mg{\small II} virial masses for the brightest quasars at each redshift is that corrections might not be necessary, and if necessary, are likely substantially smaller than previously proposed.  Most of the RYE09 correction may indeed be a result of measurement difficulties for Mg{\small II}, which become unimportant for the brightest quasars with the best-measured spectra.

\section{Discussion}
\label{sec:discussion}

The rapid falloff in quasar number density at fixed luminosity near the sub-Eddington boundary (SEB),
places a strong upper bound on the statistical uncertainty of virial mass
estimation, of, on average, $0.21$ dex for H$\beta$-based virial masses, $0.15$
dex for Mg{\small II}, and $0.27$ dex for C{\small IV}.
We considered the possibility that the narrow spreads were induced by quasar
variability.  
We find that variability can explain only $\sim$0.15~dex of spread.

The small scatter in Mg{\small II}-based virial masses at the SEB implies, surprisingly, that the Mg{\small II} masses are more reliable
than the H$\beta$-based masses. We also find that, at least for the most
luminous objects in each redshift bin, any Mg{\small II} to H$\beta$ mass corrections are likely smaller
than previously thought.  

We note that the ordering of the decay sizes is inverse to the ordering of the
emission lines by distance from the ionising continuum, as derived from
reverberation mapping \cite{Peterson2004}.
This ordering suggests a physical explanation in which the non-virial motions in the broad-line
region become relatively weaker with increasing distance from the continuum.
An obvious candidate is a reduced role of radiation pressure at larger distances
\cite{Marconi2009}, which would affect C{\small IV} most strongly and
Mg{\small II} the least. 

The measured decays provide a fixed error budget to be distributed between non-virial motion and quasar luminosity variability.  If fluctuations in the adjacent continuum scale linearly with the quasar bolometric luminosity, the $\sim 0.15$~dex bound on Mg{\small II} masses limits quasar variability to $\sim 0.33$~dex.  If the decay rate is due primarily to non-virial motion in the broad line region, then the contribution from quasar luminosity variability is smaller.  

However, if the statistical uncertainty is truly smaller in virial masses than previously believed, then
we must explain the origin of the $\sim 0.4$~dex scatter between virial and
reverberation masses, which would have to be dominated by uncertainties in
reverberation masses.  This situation is not easy to understand.  It would be remarkable if the continuum fit at one wavelength from a single-epoch spectrum (virial masses) provides a more precise radius indicator than a time-series of high-resolution spectra (reverberation masses).  
Both virial mass estimation and reverberation mapping are based on assumptions
about quasar geometry, so some of these assumptions would need to be wrong.
The basic difference between reverberation and virial masses lies in how
each method determines the radius to broad emission lines, so the larger statistical uncertainty may lie in using time delays to estimate the radius.  

Reverberation data is also used to calibrate the relation that allows the use of adjacent continuum as a proxy for radius in virial masses.  A restriction to the best reverberation data includes scatter corresponding to typical uncertainties of $0.09$~dex for broad line region radii from reverberation mapping, with an $0.11$~dex uncertainty for the H$\beta$ line \cite{Petersonpriv}.  If the scatter in the $R-L$ relation is uncorrelated with other errors, this would restrict the remaining uncertainty to just $\sim 0.1$~dex for Mg{\small II}.  Further, these uncertainty measurements use an improved $R-L$ relation \cite{Bentz2009}, while the Shen et al. (2008)\nocite{Shen2008} catalogue does not include these improvements.

The two methods also differ in that virial estimates use the entire emission line,
while the best reverberation methods use the rms line profile, i.e., the
minority fraction of the emission line that responds to continuum
changes.  Korista et al. (2001)\nocite{Korista2001} show that only the gas that is optimally emitting
a given line will respond to continuum changes.  The virial method averages
over gas in the entire broad line region.  In some manner, virial masses appear to be giving better values than reverberation masses.  We cannot satisfactorily explain this surprising result.

\section{Conclusions}
\label{sec:conclusions}

We used the sub-Eddington boundary (SEB) in the quasar mass-luminosity plane to
compare different virial mass estimators for quasar black hole masses.  In
particular, the Mg{\small II} estimator developed by McLure \& Dunlop
(2004)\nocite{McLure2004} has been considered less reliable than the
H$\beta$-based estimator of Vestergaard \& Peterson
(2006)\nocite{Vestergaard2006}, with several proposals for corrections
\cite{Onken2008,Risaliti2009,Marconi2009}.
The quasar $M-L$ plane indicates, surprisingly, that Mg{\small II} may
be the most reliable virial mass estimator.  A decline in the relative importance of non-virial motions at large radii may account for the differences in precision when using different emission lines.

Surprisingly, using the adjacent continuum as a proxy for radius seems
to be more precise than using time delays.
If the statistical uncertainties of virial mass estimates are really smaller
than previously believed, this would be a substantial improvement.  It means
that we produce better black hole mass determinations using one lower-resolution
optical spectrum than from a time-series of high-resolution spectra.

Moreover, many of the conclusions drawn from quasar mass functions, in Papers I and II, and in other
work rely on the ability to segregate quasars into mass bins.  With an
uncertainty of 0.4~dex, cross-contamination is a concern.  An
uncertainty of 0.15~dex allows us to divide a quasar sample into 2-3 times as many independent
bins.  Given the large systematic uncertainties in fitting bolometric
luminosities to templates, it is possible that quasar mass functions will be
more reliable than luminosity functions.

The quasars that define the SEB are necessarily the closest to the
Eddington limit of their cohort.  It could be that this gives them a
greater uniformity of properties, including more accurately virial
motion in the broad-line region, than quasars at lower Eddington rates. Lower Eddington rate
quasars, such as those used in reverberation mapping, might then have a
wider dispersion in these same properties, leading to the difference in observed spread.  However, since the location SEB moves with redshift, something more complex than just the Eddington ratio would need to be responsible for this possible greater uniformity in quasar properties near the SEB.

We also compared the SEB slopes using Mg{\small II} masses and H$\beta$.
Proposed corrections to Mg{\small II} masses comparing the two methods for the
entire SDSS catalogue are quite severe, but these corrections might be due to
uncertainties in the measurement of Mg{\small II} lines parameters.  For the
brightest objects at $0.4 < z < 0.8$, the SEB produced by the Risaliti, Young,
\& Elvis (2009) correction is a significantly worse match for the SEB produced
by H$\beta$ than using uncorrected Mg{\small II}-based virial masses.

The authors would like to thank Bradley Peterson, Yue Shen, Michael Strauss, and Jonathan Trump for valuable comments.  This work was supported in part by Chandra grant number G07-8136A.

\end{document}